\documentclass{aa501}
\usepackage{graphicx}
\begin{document}
   \title{Another bar in the Bulge.}
    \subtitle{}

   \author{C. Alard
       \inst{1,2}
      }

   \offprints{C. Alard}

\institute{Institut d'Astrophysique de Paris, 98bis
 boulevard Arago, 75014 Paris France. \\
%  \email{alard@iap.fr}
%
 \and
  Observatoire de Paris, 77 avenue Denfert Rochereau,
  75014 Paris France.
 }
 \date{}
 \abstract{A map of the projected density of the old stellar population
 of the Galactic Bulge region is reconstructed using 2MASS data. 
 By making
 a combination of the H and K photometric bands 
, it is possible to overcome the effect of reddening,
 and thus penetrate the inner structure of the Galactic Bulge.
 The main structure in the map corresponds to the well documented peanut shaped bar  which is formed by the inner parts of the Galactic disk as a result of
 dynamical instabilities. 
 As suggested by numerical simulations, the projected 
 Z profile of the bar, has
 an almost exponential shape. After subtracting the 
 exponential profile associated with
 the bar, a large residual appear near the Galactic Center. This
 residual is elongated and asymmetrical, which suggest a bar structure.
 Thus we arrive at the conclusion that in addition to the main bar
 a smaller bar with a different orientation may exist in the 
 central region of the Milky Way.
 This finding makes the Milky Way very similar to a large number of 
 barred spiral Galaxies which show as well a smaller bar
 in their central regions.}
    \maketitle
\section{Introduction}
The recent release by the 2MASS collaboration of a catalogue of point
 sources covering a large fraction of the sky is of great interest for
 the study of the Galactic Structure. Since the 2MASS infrared magnitudes
 by 2MASS are insensitive to extinction, these data 
 are an ideal tool to probe the structure of the Galaxy within
 a few degrees from its center. 
 The study of the Galactic Bulge in low extinction windows situated
 at larger distances from the center revealed that the Bulge was flattened
 and asymmetric.
 This elongated bulge with marked tri$-$axial
 structure is usually called the ``bar'' . The major axis 
 of the bar is pointing towards
 the first Galactic quadrant. The presence of the
 bar is very noticeable in the stellar
 component. Stanek (1994) detected an offset in
 magnitude for clump giants on different sides of the bar. A 
 Similar asymmetry is visible in the COBE image of the
 Bulge region taken in the near infrared (Dwek {\it et al}, 1995).
 But the bar is also quite visible in the 
 pattern speed of the gas (Binney {\it et al}, 1991). 
The nature of the Bar$/$Bulge system and its formation process are not
 completely clear. 
 Spiral Galaxies like the Milky way have exponential bulges 
 (Carollo {\it et al.} 2001,  Mollenhoff, C., Heidt, J., 2000, 
 Van der Marel, Roeland P., 2000), since
 exponential bulges can be formed easily has a result of disk instability
 in the central regions, we may infer that the exponential
 bulges are formed by a dynamical perturbation of the central disk. This
 bar structure seems to dominate the central region of our Galaxy, but
 a fundamental question is to know whether other components do exist
 at smaller scales. One may expect to find some parts of the bulge that
 were not formed dynamically from disk stars, but from a
 distinct stellar population. This population should be dominant
 in the central degree, and should be detectable using the 2MASS
 data.
% Another contribution to the star counts in the central
% region may come from the spheroidal halo of the Milky Way. The tracers of the
% spheroidal halo like the RR Lyrae's shows that in a broad range of distances
% from the Galactic center, the density law is close to $r^{-3.5}$ (Wetterer \& McGraw, 1996) . This density law is observed as far as 30 Kpc from the center and as close
% as Baade's window ($\simeq$ 0.5 Kpc). Provided that the core radius
% is small, the contribution of the central spheroid might become noticeable
% in the central degree.
%
%
\section{Data analysis}
 \subsection{Basic processing}
%The data were obtained by FTP via the 2MASS survey site ({\it http://pegasus.astro.umass.edu}).
 The 2MASS catalogues  corresponding to the 
 range $|b_{II}|<10$ and $|l_{II}|<15$ were obtained from the 2MASS
 public release. In this release the coverage of this coordinate
 range is not complete ($\simeq 60 \%$), but is sufficient for
 studies of Galactic structure.
 %is not complete but is sufficient ($\simeq 60 \%$) to reveal the 
 %structure of the Galactic bulge. 
 The first step is to select the sources
 with sufficient quality (read\_flag$>$1). Using this criterion there are
 about 30 millions stars in the region of interest. Within this area
 there are a number of regions which are affected by the presence of
 bright stars and their diffraction spikes.
 It is easy to identify these areas since the star counts in the
 neighborhood are much lower than average. By making 
 star counts in a box of 0.15 sq. degree all over the frame,
 we obtain an image where the regions occupied by bright stars appear
 as dark patches. In case the counts in a box are less than 10, the relevant
 pixel in the image is considered to belong to a dark patch. Even in the lower
 density regions, the mean star counts are about 10 times larger, thus our
 cut-off does not induce any artifact. To remove any contamination from
 the dark spots to the nearby pixels, all the pixels belonging to a
 3$\times$3 mesh were also flagged.
\subsection{Canceling the effect of extinction.}
 By combining 2 photometric bands it is easy to derive a magnitude that is independent
 of
 reddening. For instance using H and K, we can derive the apparent magnitude 
$m_e$:
$$
 m_e = K - \frac{A_K}{A_H-A_K} \ (H-K)
$$
 Following  Rieke and Lebofsky (1985) we take: 
 $$
 \frac{A_K}{A_V} = 0.112 \ \ \ {\rm and}  \ \ \  \frac{A_H}{A_V} = 0.175 \ \ \ \
 \ \ \ \  
 $$ 
Leading finally to:
\begin{equation}
 m_e = K - C_K \ (H-K) \\ {\rm with:} \ C_K=1.77
\end{equation}

%We will use this reddening independent magnitude for the star counts
% presented in the continuation of this paper. This filter
% combination is very insensitive to reddening, and provided
% that we make star counts of bright stars only, the effect
% of extinction on the counts should be very limited. 
%
  Note that since the reddening law in the infrared
 is unique (Cardelli {\it et al.} 1989), the definition given in eq (1)
 is valid for any line of sight, and not only for the particular
 area probed by Rieke \& Lebofski (1985). Thus in this band the cumulative
 counts of the stars brighter than some cut$-$off in magnitude $M_C$ 
 will be independent
 of the reddening (provided that $M_C$ is brighter than the observational 
 limiting magnitude). Considering that the limiting magnitude in the H band
 is about 16 and that the maximum extinction is about $A_V \simeq 30$, we
 infer that the cut$-$off should be brighter than $H=11$. Since the color
  of the tip of the upper giant branch is $H-K \simeq 0.5$ we obtain
 finally in the $m_e$ band, $M_C \simeq 9.6$. For security we
 will adopt the very conservative bright upper cut$-$off in 
 magnitude $M_C = 9$. 
  \begin{figure}[htb]
   \centering
   \includegraphics[width=7.0cm]{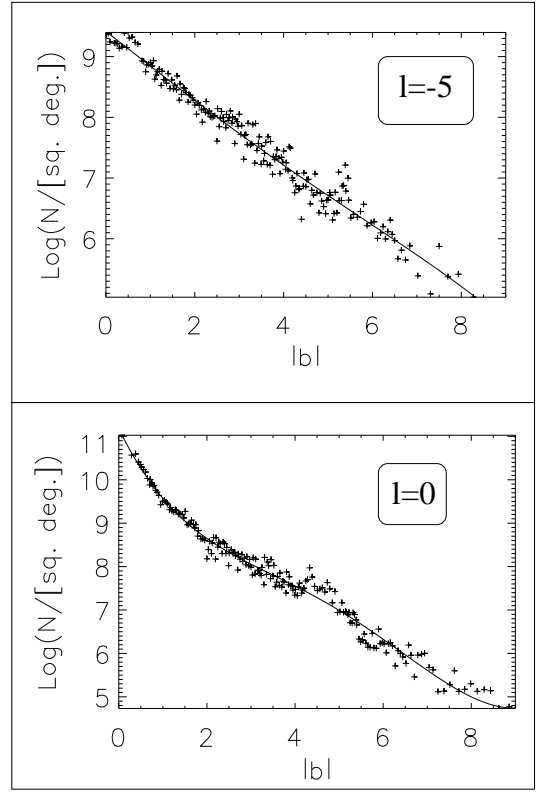}
   \caption{Here we present the polynomial smoothing of 2 sections 
 of the projected density taken at constant longitude. Note that
 the profile deviates from an exponential near the Galactic Center.}
    \end{figure}
\subsection{Reconstructing density maps of the bulge region.}
 There is one major difficulty in
  producing a map of the Bulge region: the coverage
 is not complete and additionally there are holes due to
 the bright stars. However provided that we assume that
 the Galaxy is symmetrical about its plane, it is
 possible to fill most of the gaps. In order to smooth
 and fill the smaller data gaps which remain, we use
 the fact that the density profile at constant longitude
 is almost exponential (see Fig. 1). This property suggest that
 the profile can be represented by a polynomial function.
 Numerical experiments shows that it is not significant to
 increase the degree of the polynomial beyond 5 to represent the data.
 To increase the numerical stability of the fit we will use  a strip
 of 9 columns
 centered around the column of interest. This procedure
 is carried out for each column in the image. A filtered
 image is constructed by replacing each column in the original
 image by the polynomial solution. Some example of polynomial fitting
 of the columns are given in Fig. 1. This procedure has a good ability
 to fill or extrapolate small data gaps, and has an excellent numerical
 stability. Once we have reconstructed this image of the star
 counts in the Bulge region, we apply a final wavelet smoothing procedure in
 order to balance the smoothing at all scales. This final smoothing
 is interesting because the polynomial reconstruction is one dimensional,
 and thus is biased in one direction, on a particular scale. The Wavelet
 decomposition is obtained by applying iteratively the Spline
 filter to the image and the smoothed images (Starck \& Murtagh, 1994).
 To estimate
 the statistical cuts to apply in the wavelet decomposition 
 we generate Monte-Carlo images with counts approaching our own image. 
 In the final reconstruction of the smoothed image we use 4 $\sigma$
 cut$-$off. 
 One final concern is the possible effect of
  small uncertainties in the determination
 of $C_K$ (typically a few \%). This can be investigated by reconstructing
 the density with small variations of $C_K$. The comparison
 with the initial map shows that the density variations induced by
 such changes in the value of $C_K$  are about the amplitude
 of the noise, and thus  do not affect the final result.
  \begin{figure*}
   \centering
   \includegraphics[width=11.45cm]{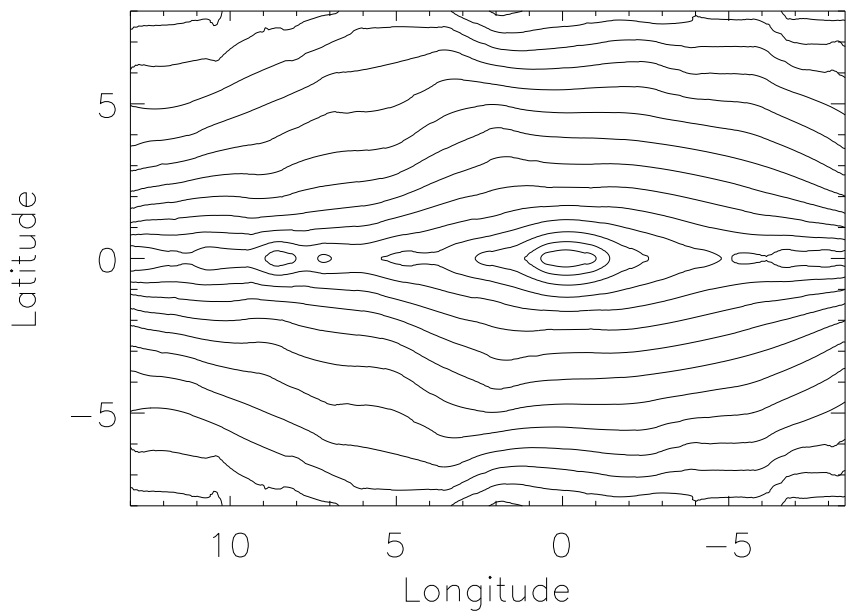}
   \caption{Map of the
  Galactic Bar region reconstructed
 using a polynomial fitting method and wavelet smoothing.
 Note that the star
 counts are systematically higher at positive longitude for $|b|>2$.
Contours values: (max,min)=(60000,400) stars/sq. deg.}
 %The value of the maximum and minimum coutour corresponds respectively to: 50000, and 3000 stars per sq. deg.}
    \end{figure*}
\subsection{Analysis of the projected density.}
We have already noticed that the Z profile of the projected density 
is almost exponential. This exponential profile is also present in numerical
 simulations of peanut shaped bars. Combes {\it et al.} (1990) showed that
 a disk with a small bulge near its center forms a peanut
 shaped bar with a nearly exponential projected Z profile (see in particular
 Fig. 4 in Combes 1990). Thus if we
 subtract the exponential contribution which corresponds to the bar$-$disk
 system, the remaining density may reveal another component. 
 The contribution of the bar will be estimated by fitting an exponential
 to each column of the image (which corresponds to the projected Z profile).
 This procedure is more flexible than trying to subtract a given bar model. 
 There are
 still many uncertainties concerning the structure
 of the Galactic bar, thus
  the subtraction of a bar model may 
 give ambiguous results. By fitting an exponential
 profile, we make no particular assumption about the shape of the bar, other than 
 a general assumption on the Z profile at equilibrium which
 is justified by numerical simulations.
 To implement the fit of the exponential profile, we perform
 a robust fitting of a straight line to the log of density 
 (by minimizing the sum of absolute deviations).
 Once the exponential contribution
 has been subtracted, a very significant residual appears
 in the central region ($R<2\hbox{$^\circ$}$). The contours of this
  residual are smooth and elongated along the Galactic plane. This
 component shows also a very significant asymmetry in longitude. The
 amplitude of this asymmetry is close to 15 \%, which is about
 7 sigmas according to Poisson statistics.
 There are also some residuals along the Galactic plane in general.
But their amplitude is about 10 times smaller, they
 do not have smooth structure, and their
 scale is much smaller. These residuals are probably due to the presence of 
young stars in the HII regions.
  \begin{figure*}
   \centering
   \includegraphics[width=11.45cm]{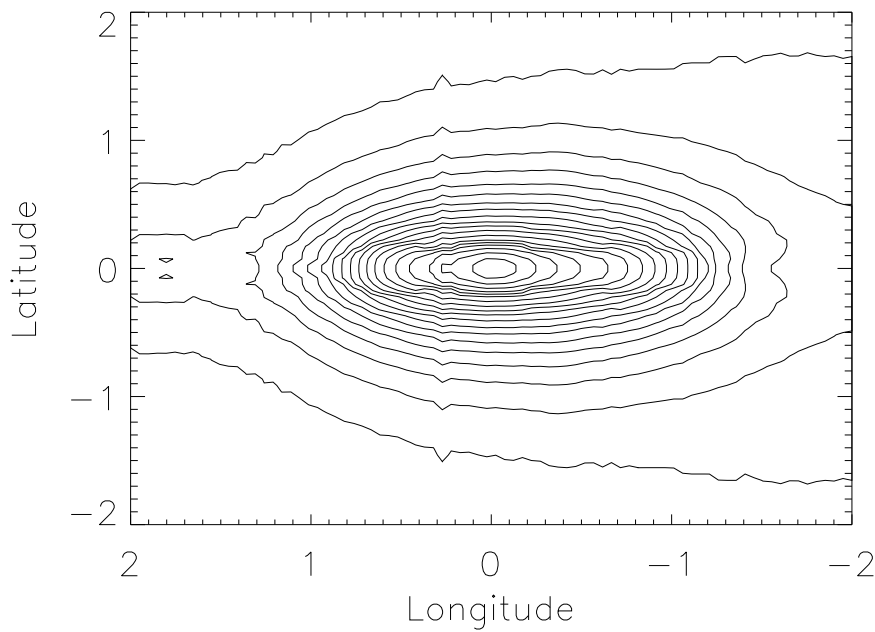}
   \vspace{0.3cm}
   \caption{Contours of the residual density after subtracting the main component.
 Note that the asymmetry is in the opposite direction of the large scale
 Galactic Bar. Contours values: (max,min)=(38000,400) stars/sq. deg.
}
    \end{figure*}
\section{Discussion}
There are some intrinsic difficulties in the interpretation of the
 former results, we see a residual component after subtracting
 a main component. But this result is based on the assumption that
 the bar has a projected density profile in the Z direction which 
 is exponential. Even if this assumption is supported by numerical
 simulations, it might just be that the projected density of the bar is
 not exponential. But in this case, why do we observe an asymmetry
 in an opposite direction to the bar ? 
 Two possible sources of bias are: the residual extinction, and
 the steepness of the density profile near the center (Blitz \& Spergel 1991).
 In principle the extinction should have no effect in the $m_E$ band,
 but it is possible that for observational reasons the limiting magnitude 
 in 2MASS is somewhat brighter than expected, which would result in
 an indirect extinction effect.
 These biases can be investigated by using numerical simulations.
  To build 
 a numerical model we need to integrate the convolution 
 product of the luminosity function $\phi$ with the density
 distribution $\rho$. The integration domain in the
 space of the magnitudes will be modulated by the extinction
 $A_V$. For the luminosity function
 we will adopt the model of Wainscoat {\it et al} (1992). The density
 distribution will be built using a truncated exponential disk (Lopez$-$
 Corredoira {\it et al.} 2001) and a
 triaxial bar model with a power law profile. This bar model has
 an inclination of 20 deg. with respect to the line of sight, and
 axis ratio: $\frac{x_0}{y_0} \simeq 3$ and $\frac{z_0}{y_0} \simeq 0.7$
 (Dwek {\it et al.} 1995). And finally the extinction map of the whole
 area was built from our 2MASS data
 by using a method presented by Schultheis {\it et al.} (2000).
 Let's start with the case of residual extinction effects: the
 counts were generated using the aforementioned procedure, Poisson noise was
 simulated,  and finally the whole process of polynomial reconstruction,
 smoothing and exponential subtraction was applied. This procedure
 was undertaken for different observational limiting magnitudes, starting
 from our default value (no extinction cut$-$off). To summarize the
 results, the longitude profile of the residual density has been
 represented for the different limiting magnitude (Fig. 4). The comparison
 of these results with the observational profile shows unambiguously
 that the effect of extinction is in the opposite direction 
 to the observed asymmetry. Furthermore, it is interesting
 to notice the steep edge of the observational profile, which
 is similar on both side of the diagram. The profile variations
 produced by extinction are not similar on both edges, and are hardly
 so steep. Concerning the effect of the steepness of the inner bar
 profile, even if a steep bar profile, may produce an asymmetry
 opposite to the bar, it is not possible 
 reproduce the amplitude of the asymmetry even if the density 
 profile behaves like: $R^{-7}$.A steeper density profile appears
 unphysical.
 These results are not very dependent upon the luminosity function.
 Power law luminosity functions gives the same result.
 The best explanation is to consider the presence
 of another component in the inner Galaxy.
 A small bar with a steeply dropping density near its edge
 can reproduce the observed asymmetry. This finding
 is not surprising, the discovery of small
 substructure in the central region of barred spiral Galaxies
 is very common (Erwin \& Sparke 1999, Fiedli {\it et al.} 1996, Friedli 1996, Shaw {\it et al.} 1995).  
 This small bar structure was
 not found in previous studies for 2 fundamental reasons,
 the former data set lacked either the depth or the resolution.
 The 2MASS H band is deep
 and insensitive to extinction, and furthermore the 30
 million stars available result in 
 an excellent spatial resolution. 
%
 %In most case these smaller structure can be interpreted
 %as a smaller bar which is rotating at a different speed. The Milky Way
 %has probably a similar system of 2 bars, with different
 %orientations, and rotation speed. Further spectroscopic observations 
 %in the infrared will help to estimate the rotation speed, velocity dispersion
 %and stellar population characteristics of this component of the Galaxy.
%
  \begin{figure}
   \centering
   \includegraphics[width=7.0cm]{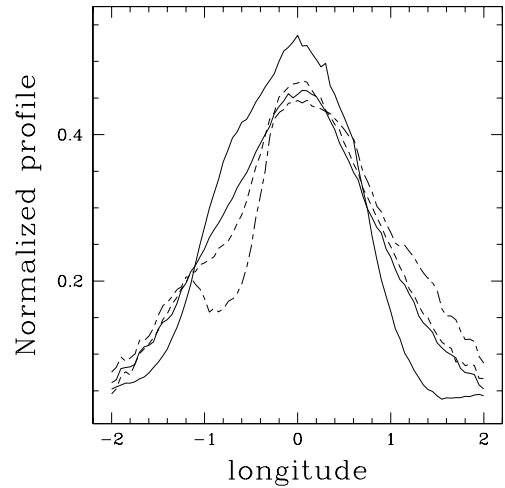}
   \caption{The marginal distribution in longitude of simulated profiles
 for different limiting magnitude (default: thin line, 1 mag. brighter
 than default: dashed line, 1.5 mag. brighter: dotted dashed line). The
 last profile is the observational profile (thick line). All profiles
 have been normalized so that the sum of the profile is unity.}
    \end{figure}
\begin{acknowledgements}
This publication makes use of data products from 2MASS, which is a 
joint project of the Univ. of Massachusetts and the Infrared Processing 
and Analysis Center, funded by the NASA and the NSF. \\
It is a pleasure to thank Alain Omont and Bohdan Paczy\'nski for
interesting discussion. I would like to thank Dave Spergel for making
an interesting point. I would like to thank Michael Friedjung for
 checking the final proof of this manuscript.
\end{acknowledgements}
\end{document}